\documentclass[%
 reprint,
 amsmath,amssymb,
 aps,nofootinbib,
]{revtex4-1}
\usepackage{graphicx}
\usepackage{dcolumn}
\usepackage{bm}
\PassOptionsToPackage{linktocpage}{hyperref}
\usepackage[hyperindex,breaklinks]{hyperref}
\usepackage{tabularx}
\usepackage{enumitem}
\usepackage{slashed}

\usepackage{array}
\usepackage{mathtools}

\usepackage[english]{babel}
\usepackage{fancyhdr}
\usepackage{setspace}
\usepackage{graphics}
\usepackage{eso-pic}
\usepackage{amsmath}
\usepackage{amstext}
\usepackage{amsfonts}
\usepackage{amssymb}
\usepackage{hyperref}
\usepackage{slashed}
\usepackage{multirow}
\usepackage{esdiff}
\usepackage{commath}
\usepackage{tikz-cd}
\usepackage{relsize}
\usepackage{scalerel}
\usepackage{enumitem}

\usepackage{tabularx}

\usepackage{mathrsfs} 
 
\tikzstyle{RectObject}=[rectangle,fill=white]

\renewcommand{\vec}[1]{\ensuremath{\boldsymbol{#1}}}

\newcommand{\iu}{\mathrm{i}}	
\newcommand{\ee}{\mathrm{e}}	

\newcommand{\Scri}{\mathscr{I}} 

\newcommand{\bra}[1]{\ensuremath{\left< #1\,\right|}}
\newcommand{\ket}[1]{\ensuremath{\left|\, #1\right>}}

\begin{document}
\begin{abstract}
\noindent
In the present note we show that the recently established connections between soft theorems, large gauge transformations and memories are persistant in the infrared safe formulation of quantum field theory. They take a different and simplified form and can all be derived from the non-trivial asymptotic dynamics that is proper to any theory with massless fields. Most of the results in this paper were already presented in \citep{CesarMischa} and, with a different interpretation, in \cite{gabai} and \cite{shavingHair}. The new parts (as compared to \cite{CesarMischa}) are an improved derivation of charges for large gauge transformations in the framework of non-trivial asymptotic dynamics, the connection to the classical memory effect and an overall more accessible treatment of the topic. Since the formulation of QFT without infrared divergences is physically more appealing, the infrared safe version of the above connections should be so as well. 
\end{abstract}

\title{The infrared triangle in the context of IR safe S matrices}
\author{Mischa Panchenko}
\affiliation{%
 Arnold Sommerfeld Center, Ludwig-Maximilians-Universit\"at, Theresienstra{\ss}e 37, 80333 M\"unchen, Germany
}%
\email{m.panchenko@campus.lmu.de}
\maketitle

\newpage

\section{Introduction}
In the last years a remarkable universal structure in classical and quantum field theories, dubbed the infrared triangle, was discovered and studied in detail, see e.g.  \cite{masslessQED, memoryStrom, LGTinQED, memoryPast} and references therein. A central result underlying these studies goes under the name of soft theorems - a relation between amplitudes in gauge theories with a soft particle in the final state to amplitudes without such particles. The soft theorems are a manifestation of a general problem in QFTs with massless particles: the amplitudes for soft emission are divergent and hence such theories are said to possess no S matrix. How can it be that a manifestation of the problematic infrared divergences is related (and in fact equivalent) to fundamental symmetries and even to observable effects such as memories? In order to answer this natural question one needs to first focus on the resolution of the IR problem in QFT and then see how the triangle manifests itself in the divergence free, IR safe theory.  In the present paper we will do just that. Since in the IR safe theory the soft photon theorem is replaced by decoupling of soft particles, the triangle becomes in a sense even simpler than before.\\

Luckily, most of the work was already done (and in fact done long ago) so all that is left is to connect the dots and make the picture apparent. To keep the paper short, we will refer the reader to original papers and a recent longer paper \cite{CesarMischa} co-authored by the author himself for details. \\

The key point of the present paper is the well known realization that in the presence of massless particles, i.e. long range interactions, the early and late time dynamics cannot be treated as free. Instead, non-trivial asymptotic dynamics have to be found and implemented in the construction of the S matrix. Once this is done, the S matrix is free from IR divergences. As we will show, the three corners of the infrared triangle are all consequences of these asymptotic dynamics. They are

\begin{enumerate}
  \item \textbf{Soft theorem}: Soft particles decouple from the IR safe S matrix 
  \begin{equation}
 \lim_{\omega\rightarrow 0} [\omega\, a^\lambda(\omega\vec{e}_k) , S]=0 
  \end{equation}
 
  \item \textbf{Large gauge transformations}: The charge of a LGT consists purely of soft particles, the non-trivial asymptotic dynamics kills the hard part. It thus commutes with the S matrix and this decoupling implies antipodal matching.
  
  \item \textbf{Memories:} The asymptotic value of the (zero frequency of the) massless field (scalar, photon or graviton) carries memory about the scattering process. This memory has a classical manifestation.
\end{enumerate}

The picture of the triangle hence gets modified to a tetrahedron. For clarity and to be concrete, throughout the paper we will focus on QED, but the discussion can be straightforwardly  carried out for other field theories with massless particles, like gravity. 
\begin{center}

\begin{figure}\label{pyramid}
\begin{tikzpicture}
\draw (0,0) node[anchor=south](A){\textit{Decoupling of soft modes}}
  -- (3,-3) node[anchor=north](C){\textit{Memories}}
  -- (-3,-3) node[anchor=north](B){\textit{LGT and matching}}
  -- (0,-1.5) node[RectObject] (D){\textit{As. dynamics}}
  -- cycle;
  \draw (A.south)--(B.north);
  \draw (D)--(C.north);
\end{tikzpicture}
\caption{Infrared tetrahedron}
\end{figure}
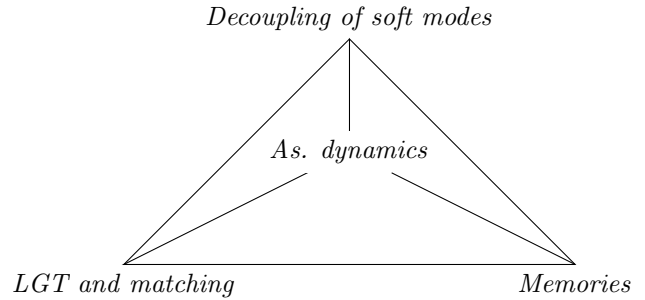
\end{center}

\section{The triangle from asymptotic dynamics}
The first paper describing the modification of the asymptotic dynamics in presence of long range interactions was \cite{dollard}. The idea was then applied to QED in \cite{kulish, zwanziger1, zwanziger2} and many subsequent papers, see also \cite{rohrlich} for a pedagogical introduction. As established there, asymptotic dynamics in QED is governed by the following evolution operator (here and throughout the paper we use the same notation as in the corresponding references unless stated otherwise):
\begin{equation}\label{asdym}
U_{as}(t)=\ee^{-\iu H_0 t} \ee^{R(t)} \ee^{\iu \Phi(t)}
\end{equation}
with 
\begin{align}\label{R}
&R(t)=\frac{e}{(2\pi)^3}\int \frac{p^\mu}{p\cdot q}\,\rho(p)\left(a^\dagger_\mu(q)\ee^{\iu\frac{q\cdot p t}{p_0}}-h.c.\right)\frac{d^3q}{2 \omega} d^3p \\
& \Phi(t) \sim \int : \rho(p)\rho(k) : \frac{p\cdot k}{((p\cdot k)^2-m^4)^\frac{1}{2}} \text{sign}(t) \ln (\,\abs{t}) \,d^3p\,d^3k \nonumber
\end{align}
The phase operator plays no role in the following discussion and we will ignore it from now on. \\

This evolution operator gives the asymptotic current and electromagnetic field operators respectively as:
\begin{align}
J_\mu^{as}(x) & =\int d^4p\, p_\mu \,\rho(p) \int\limits_{-\infty}^\infty d\tau\,\delta^4(x-p\tau)  \label{jas}\\
A_\mu^{as}(x) & =A_\mu^{free}(x)+\int d^4y\, \Delta^{ret}(x-y) J_\mu^{as}(y) \label{Aas}
\end{align}
The expressions are understood to have physical meaning only for large $\abs{t}$ and large separations. The above expression for $J_\mu$ is simply the current of particles flying on straight lines through the origin. These equations are all that is needed to derive each corner of the triangle.

\subsection{The soft decoupling}
The definition of $U_{as}$ results the following IR safe $S$ matrix
\begin{align}
S_{KF}&=\lim_{t_\pm\rightarrow\pm\infty} \ee^{R(t_+)^\dagger}\, S_D(t_+,t_-) \ee^{R(t_-)} \nonumber \\
&=: \lim_{t_\pm\rightarrow\pm\infty} S_{KF}(t_+,t_-)
\end{align} 
Here $KF$ stands for Kulish-Faddeev and $S_D$ is the standard Dyson $S$ matrix in the free interaction picture, i.e. $S_D(t_+,t_-)=U_I(t_+,t_-)$.\\

The normal leading order soft photon theorem implies that 
\begin{equation}\label{decoupling1}
\lim_{\omega\rightarrow 0} [\omega\,a^\mu(q),S_{KF}]\, = 0 
\end{equation}
This was already stated in \cite{kulish} (though without proof).	The proof involves a careful ordering of limits. First note that
\begin{widetext}
\begin{align*}
[a^\mu(q),S_{KF}(t_+,t_-)]=  [a^\mu(q),\ee^{R(t_+)^\dagger}]\, S_D(t_+,t_-) \,\ee^{R(t_-)} &+ \ee^{R(t_+)^\dagger}\,[a^\mu(q), S_D(t_+,t_-)] \,\ee^{R(t_-)}  \\
  & + \ee^{R(t_+)^\dagger}\, S_D(t_+,t_-) \,[a^\mu(q),\ee^{R(t_-)}] 
\end{align*}
\end{widetext}
We can now use standard formulas for displacement operators and take the $\omega\rightarrow0$ limit before the $t_\pm\rightarrow\pm\infty$ limit finding (schematically):
\begin{align*}
&\quad\lim_{\omega\rightarrow 0} [\omega\,a^\mu(q),S_{KF}]  =  \ee^{R^\dagger(\infty)}  \cdot \nonumber\\
\lim_{\omega\rightarrow 0} \, \omega\, &\left( e \int \frac{p^\mu}{p\cdot q}\,[-\rho(p),S_D] dp+[a^\mu(q),S_D] \right) \ee^{R(-\infty)}
\end{align*}
The soft photon theorem is equivalent to the vanishing of the expression in-between the dressing operators and hence to equation (\ref{decoupling1}). This version of the proof was also presented in \cite{CesarMischa}. It is important to notice that the limits  $\omega\rightarrow 0$ and $t_\pm\rightarrow \pm\infty$ do not commute.

\subsection{Charges of LGT from non-trivial dynamics}
Under the assumption of free asymptotic dynamics a basis for the charges of large gauge transformations on $\Scri^+$ in massless QED was found explicitly by Strominger et al to be (in a combination of the  notation in \cite{masslessQED} and the notation in \ref{R})
\begin{widetext}
\begin{align}
Q_{free}^+(\vec{e}_x)&= \frac{1}{\sqrt{2}e}\frac{1}{1+z\bar{z}} \left( \lim_{\omega\rightarrow 0^+} \left[\omega a_{+\,free}^{out}(\omega\vec{e}_x)+h.c.\right]-2e\int \frac{\omega\,p\cdot \varepsilon^+(\vec{e}_x)}{  p\cdot q}\,\rho^{out}(p) d^3p \right)\\
&=: Q^{+ \, soft}_{free}(\vec{e}_x) + Q^{+ \, hard}(\vec{e}_x)
\end{align}
\end{widetext}
Here the notation means:
\begin{equation}
q=\omega(1,\vec{e}_x) \quad,\quad \vec{e}_x=\vec{e}_x(z,\bar{z})
\end{equation}
and $Q_{free}^+(\vec{e}_x)$ is what was called 
\begin{equation}\label{epsilon}
Q^+(\varepsilon) \quad \text{ with } \quad \varepsilon(w,\bar{w})=\frac{1}{z(\vec{e}_x)-w}
\end{equation}
in \cite{masslessQED}.\\
The relation between any expressions on $\Scri^+$ derived using free dynamics and using the full asymptotic dynamics \ref{asdym}, in particular for the charges of LGT, is
\begin{equation}
Q^+=\lim_{t\rightarrow \infty} \ee^{R^\dagger(t)} Q_{free}^+  \ee^{R(t)}
 \end{equation} 
We have not put the label ``free" on the hard part because due to the specific form of $R(t)$ it coincides with the true asymptotic operator. Using that and taking the limit $\omega\rightarrow 0$ before  $t\rightarrow \infty$ we obtain
\begin{widetext}
\begin{align*}
Q^+(\vec{e}_x)=\frac{1}{\sqrt{2}e}\frac{1}{1+z\bar{z}}\left( \lim_{\omega\rightarrow 0^+} \left[\omega a_{+}^{out}(\omega\vec{e}_x)+h.c.\right] 
+ 2e\int \frac{\omega\,p\cdot \varepsilon^+(\vec{e}_x)}{  p\cdot q}\,\rho^{out}(p) d^3p-2e\int \frac{\omega\,p\cdot \varepsilon^+(\vec{e}_x)}{  p\cdot q}\,\rho^{out}(p) d^3p \right)
\end{align*}
\end{widetext}
In other words, the non-trivial asymptotic dynamics has killed the hard part of the charge that creates LGT and we have
\begin{equation}\label{Qplus}
Q^+(\vec{e}_x)=\frac{1}{\sqrt{2}e}\frac{1}{1+z\bar{z}} \lim_{\omega\rightarrow 0^+} \left[\omega a^{out}_+(\omega\vec{e}_x)+h.c.\right] 
\end{equation}
A similar calculation applies to $\Scri^-$ and results in
\begin{equation}\label{Qminus}
Q^-(\vec{e}_x)=\frac{1}{\sqrt{2}e}\frac{1}{1+z\bar{z}} \lim_{\omega\rightarrow 0^+} \left[\omega a^{in}_+(\omega\vec{e}_x)+h.c.\right] ,
\end{equation}
where $z,\bar{z}$ are now coordinates on $\Scri^-$. Similar statements (but with different derivations and interpretations) can be found in \cite{gabai, CesarMischa, shavingHair}. The vacuum in QED is degenerate, two vacua are related to each other by LGT or equivalently by soft photons and the soft photon, being the Goldstone boson of a spontaneously broken global symmetry, decouples from the S matrix. The picture becomes very unified and simple.\\

We find it necessary to comment on the meaning of equations \ref{Qplus} and \ref{Qminus} in order to avoid possible confusion. From the fact that the charges of LGT do not contain a hard part \textit{does not follow} that they commute with the charged fields. After all, if they did, they would not create large gauge transformations anymore. There are two ways to see that expressions $Q^+$ and $Q^-$ still generate LGT despite their purely soft appearance. The simplest way is to note that when expressed in terms of asymptotic field operators, the charges are given by the expressions (3.2) and (5.6) in \cite{masslessQED} (with the appropriate $\varepsilon$, as in \ref{epsilon} ) and hence they obviously generate the correct transformations. An alternative way would be to compute the action of  $Q^\pm$ on the asymptotic expressions for field operators of charged particles. This will be done in the appendix.

\subsection{From decoupling to symmetry and matching}
Due to the decoupling of soft modes from the IR safe S matrix, \ref{decoupling1}, and the fact that charges of  LGT contain purely soft modes, it is clear that LGT constitute symmetries of the S matrix. The antipodal matching from $\Scri^-$ to $\Scri^+$ is obtained using that for any in-operator $O_{in}$
\begin{equation}
S^\dagger O_{in} S = O_{out}
\end{equation}
and from the decoupling
\begin{equation}
[S,Q^{-}(\vec{e}_x)]=0
\end{equation}
which results in the antipodal matching
\begin{equation}
Q^{-}(\vec{e}_x(z,\bar{z}))=Q^{+}(\vec{e}_x(z,\bar{z})).
\end{equation}
See also \cite{CesarMischa} for an extended discussion of this topic.

\subsection{From asymptotic dynamics to memory}
The connection of the topics treated above to classical memory is surprisingly simple and is most easily understood by comparing the works of Zwanzinger \cite{zwanziger1}  and Rohrlich \cite{rohrlich} to the recent work by Tolish, Wald et al. \cite{tolish1, tolish2}. The crux is in the equations \ref{jas}, \ref{Aas} which appear in both approaches. A classical scattering process from particles with incoming momenta $\{p^{in}_i\}$ to $\{p^{out}_j\}$ can be understood within QFT as having the system in the state $\ket{p^{in}_i}$ in the far past and in $\ket{p^{out}_j}$ in the far future. Assuming the interaction is to happen instantaneously at time $t=0$, as it was done in \cite{tolish1,tolish2}, it is appropriate to use the asymptotic expressions \ref{jas}, \ref{Aas}  for all times $t\neq 0$. The momenta eigenstates $\ket{p^{in/out}}$ are eigenstates of the asymptotic current, hence assuming that the state $\ket{p^{in}_i}$ evolves into the state $\ket{p^{out}_j}$ at time $t=0$ and computing the expectation value of $J^{as}_{\mu}(\vec{x})$ results in the current given by formulas (21) and (22) in \cite{tolish2}. Then the expectation value of the asymptotic electromagnetic potential $A^{as}_{\mu}(x)$ coincides with formula (24) in \cite{tolish2}, which leads to electromagnetic memory. \\

One can also make a direct connection to memories without referring to the work by Tolish et al. As described in \cite{memoryBieri} and \cite{memoryPast}, the classical EM memory (in the absence of charged massless particles)  can be reconstructed from 
\begin{equation}
\Delta\,W=\lim_{r\rightarrow\infty}  r\,A_{u}(u=\infty,r,\vec{e}_x)-r\,A_{u}(u=-\infty,r,\vec{e}_x)
\end{equation}
where we rewrote equation (46) from \cite{memoryBieri} in terms of $\Scri^+$ quantities, just as it was done in \cite{memoryPast}. Now in QFT one must replace  $A_{u}$ by $\left< A_u\right>$. In the far past and future the time evolution is given by the formulas \ref{jas} and \ref{Aas}. Thus, one can find the memory of a scattering process by computing
\begin{equation}\label{Aaout}
\lim_{u\rightarrow\infty} \bra{p^{out}_j}\lim_{r\rightarrow\infty}  r A^{as}_u(u,r,\vec{e}_x)\ket{p^{out}_j}
\end{equation}
and
\begin{equation}\label{Ain}
\lim_{u\rightarrow-\infty} \bra{p^{in}_j}\lim_{r\rightarrow\infty}  r A^{as}_u(u,r,\vec{e}_x)\ket{p^{in}_j}
\end{equation}
and using equations \ref{jas} and \ref{Aas} or equivalently directly \ref{asdym} and the usual identities for coherent states. In that context see also \cite{dressedCharges, zwanziger3} and references therein. This results precisely in the formula (48) from \cite{memoryBieri}. \footnote{That can be seen either from direct computation or from the form of $J^{as}_\mu$ and a comparison to the classical treatment of \cite{landau} or also from the treatment of \cite{tolish2}} In this context see also \citep{CesarRaoul} where memories were studied from a different point of view.
\subsection{Memories and large gauge transformations}
In classical electromagnetism the angular components of the field $A_{\mu}$ are pure gauge (i.e. derivatives of a scalar) at $u\rightarrow\pm\infty$ for any scattering and hence the electromagnetic memory is found from the difference of two pure gauge configurations at $\Scri^+$ - a large gauge transformation. This can be immediately read off the formula (3.3) in \cite{memoryPast} which coincides with the fixed angle LGT charge (3.4) of \cite{memoryStrom}. At the same time we found that soft photons and hence the charges of LGT decouple from the $S$ matrix. Note that these statements are not in conflict with each other - also in the previous treatments charges for LGT decoupled from the (IR divergent) S matrix. However, since in the IR safe scenario the probability to excite soft photons is zero - just like the probability to tunnel from one vacuum to another in any field theory with a continuous vacuum degeneracy - one cannot say that the physically real memory effect is directly related to an emission of soft particles. The soft factor from the scattering amplitudes determines the memory of a scattering process because it enters the dressing operator $R(t)$ and this in turn is the only important quantity for the evaluation of \ref{Aaout} and \ref{Ain}. This completes the infrared tetrahedron (figure \ref{pyramid}) in the context of IR safe theories.   

\subsection{A note on other field theories}
Although we have focused on electrodynamics in this paper, most of the derivations and concepts hold for any theory with massless particles. For example, the derivation of the infrared triangle in  gravity is a straightforward (though technically more challenging) analogy. However, the triangle seems incomplete for the case of massless scalars with the missing corner being large gauge transformations. All other corners are still there for scalars - the asymptotic dynamics is nontrivial, there is a coherent dressing operator $R(t)$, the soft decoupling and also a memory effect (see \cite{tolish2} for the latter). However, the soft decoupling does not seem to correspond to any classical symmetry like the LGT in gauge theories. It might be possible to uncover large gauge transformations for scalar fields by using their duality to gauge fields, see \cite{duality} \footnote{This idea was communicated to me by Gia Dvali}.

\section{Comparison with the triangle in the IR diverging theory}
Just as in the above references, the related literature on the infrared triangle usually considers the asymptotic dynamics for field operators to be free.  In such a setup the antipodal matching of LGT is equivalent to the usual soft theorems which encode infrared divergences of the quantum theories. Thus, it might appear that the divergences of amplitudes follow from fundamental symmetries and hence cannot be circumvented. This would be particularly inconvenient, since usually in the very same treatments  the notion of the S matrix is used - but it is precisely the IR divergences that lead to the non-existence of such an operator. The connection to memory is then done through analogues of formula (5.6) of \cite{memoryStrom} where the ratio of two amplitudes is interpreted as the asymptotic expectation value of a field operator. To the author's knowledge there is no a priori derivation of this relation. \\

In the IR safe scenario the picture is significantly clearer. An IR safe S-matrix exists and soft particles decouple from 	it. \footnote{This makes immediate physical sense since the time scale on which soft particles interact is infinity.} LGT are generated by purely soft quanta and their antipodal matching is equivalent to the soft decoupling. The memory effect is found directly from studying asymptotic expectation values for field operators for a determined scattering process, no formulas like the above mentioned (5.6) of \cite{memoryStrom} are needed. \footnote{Note that from the soft decoupling follows that the ratio of amplitudes in such expressions vanishes.} 
\section{Conclusions and outlook}
In this paper we have presented the infrared triangle within the IR safe theory where it is enhanced to a tetrahedron. All corners and the connections between them have been highlighted. A natural next step would be to include subleading (in the frequency of the soft particles) results. In order to do that, subleading terms in the dressing operators $R(t)$ would be needed to be found. A first step in this direction was taken in \cite{CesarMischa}, but the construction there was rather ad hoc. A more natural expansion in frequency would be preferable. Furthermore, the IR safe S matrix has not been made into a practical tool yet, in particular, there is no simple diagrammar for it (although some attempts and calculations have been carried out - see \cite{zwanziger3}, \cite{masterInfra} and references therein). It would also be very interesting to find the manifestation of the soft decoupling for other space-times different from Minkowski, in particular, for space-times with horizons where due to infinite redshift the infrared structure is much richer. We plan to address some of these issues in future work.

\section{Appendix}
It is clear that the charges \ref{Qplus} and \ref{Qminus} generate the correct gauge transformations on the asymptotic operator $A_{\mu}^{as}$ but less obvious that they also generate the right transformation for matter fields. In order to explicitly demonstrate that on an example, we look at the asymptotics of a massless scalar field $\phi$ on $\Scri^+$. Using the saddle point approximation for large $r$  (see \cite{LGTinQED} and the appendix of \cite{CesarMischa}) and the asymptotic time evolution \ref{asdym} one obtains
\begin{equation}
 \phi(u,r\gg 1, \vec{e}_x)\sim \frac{1}{r}\, \ee^{R(\vec{e}_x,t)}  \int  \,\ee^{\,\iu\,u\,p}\,\, b_{out}^\dagger(p,\vec{e}_x) \,dp \, 
\end{equation} 
where we only write out the particle but not the anti-particle part and we have defined
\begin{equation}
R(\vec{e}_x,t):=\frac{e}{(2\pi)^3}\int \frac{p^\mu}{p\cdot q}\,\rho(p)\left(a^\dagger_\mu(q)\ee^{\iu\frac{q\cdot p t}{p_0}}-h.c.\right)\frac{d^3q}{2 \omega} 
\end{equation}
with
\begin{equation}
p_\mu=p_0(1,\vec{e}_x)
\end{equation}
so that the above expression is indeed independent of $p_0$. Then using the commutation relation

\begin{equation}
\lim_{\omega\rightarrow 0}[\omega a_\pm(\omega \vec{e}_y), \ee^{R(\vec{e}_x,t)} ]= e  \frac{p\cdot \varepsilon_\pm(\vec{e}_y)}{p\cdot(1,\vec{e}_y)}\ee^{R(\vec{e}_x,t)}
\end{equation}

one obtains that on $\Scri^+$
\begin{equation}
[Q^+(\vec{e}_y),\phi(u,\vec{e}_x)]=-\varepsilon(\vec{e}_y) \phi(u,\vec{e}_x)
\end{equation}

as it must be. Here we again used the notation as indicated in \ref{epsilon}.
\section{Acknowledgments} 
I would like to thank Cesar Gomez for many useful discussions and for bringing several important references to my attention.
This work was supported by the ERC Advanced Grant ``UV-completion through Bose-Einstein Condensation" (Grant No. 339169). 

\newpage

\bibliographystyle{apsrev4-1}
\interlinepenalty=10000
\bibliography{bibfile}

\end{document}